\documentclass[prl,aps,twocolumn,showpacs,amsmath,amssymb]{revtex4}
\usepackage{epsfig}

\def\Sr327{Sr$_3$Ru$_2$O$_7$}

\begin{document}

\title{Thermal conductivity in the vicinity of the quantum
critical endpoint in \Sr327}

\author{F. Ronning$^{1,2}$, R.W. Hill$^{1\dagger}$, M. Sutherland$^{1\star}$,
D.G. Hawthorn$^{1\ddagger}$, M.A. Tanatar$^{1\S}$, J.
Paglione$^{1\natural}$, Louis Taillefer$^{1,3}$, M.J. Graf$^2$,
R.S. Perry$^{4,5}$, Y. Maeno$^4$, A.P. Mackenzie$^5$}

\affiliation{$^1$Department of Physics, University of Toronto, Toronto, Ontario, Canada\\
             $^2$Los Alamos National Laboratory, Los Alamos, New Mexico 87545, USA\\
             $^3$Universit\'{e} de Sherbrooke, Sherbrooke, Quebec, Canada\\
             $^4$Department of Physics, Kyoto University, Kyoto 606-8502, Japan\\
             $^5$School of Physics and Astronomy, University of St. Andrews, Fife KY16 9SS, Scotland}

\date{\today}

\begin{abstract}
Thermal conductivity of \Sr327~was measured down to 40~mK and at
magnetic fields through the quantum critical endpoint at $H_c$ =
7.85~T. A peak in the electrical resistivity as a function of
field was mimicked by the thermal resistivity. In the limit as
$T\rightarrow$ 0 K, we find that the Wiedemann-Franz law is
satisfied to within 5\% at all fields, implying that there is no
breakdown of the electron despite the destruction of the Fermi
liquid state at quantum criticality. A significant change in
disorder (from $\rho_0$($H$=0T) = 2.1 $\mu\Omega$ cm to 0.5
$\mu\Omega$ cm) does not influence our conclusions. At finite
temperatures, the temperature dependence of the Lorenz number is
consistent with ferromagnetic fluctuations causing the non-Fermi
liquid behavior as one would expect at a metamagnetic quantum
critical endpoint.
\end{abstract}
\maketitle

While classical phase transitions are theoretically well
understood, quantum phase transitions are in defiance of
theoretical understanding. At a quantum critical point (QCP), the
Fermi liquid ground state is destroyed by the diverging quantum
fluctuations associated with a particular phase transition. Given
that the physics up to very high temperatures can be dominated by
the presence of a QCP, it is essential to try and understand the
nature of the fluctuations and excitations which exist at a QCP.
Part of the problem in understanding quantum phase transitions is
that, given nearly a hundred different systems which show
non-Fermi liquid behavior presumably due to proximity to a QCP,
there is very little commonality between various observables, such
as resistivity, susceptibility, and specific
heat.\cite{StewartRMP} What is desperately needed are very
fundamental measurements of physical properties in the vicinity of
a QCP.

The Wiedemann-Franz law (WFL), which states that thermal
($\kappa$) and charge ($\sigma$) conductivities are simply related
through the expression $\kappa/\sigma T = L_0$, where $L_0$ is the
Sommerfeld value of the Lorenz number (2.44$\times$10$^{-8}$
W$\Omega$/K$^2$), is precisely such a fundamental probe of
strongly correlated physics. At $T$~=~0 the law is a consequence
of the fact that all fermionic excitations carry charge $e$, while
all of the possible bosonic excitations have zero charge. Should a
violation be expected at a QCP? Experiments on established quantum
critical systems, such as specific heat and resistivity on
YbRh$_2$Si$_2$, have been interpreted as observing the breakup of
the electron at a QCP,\cite{CustersNature2003} which would naively
imply a violation. In addition, theories of quantum criticality
are also suggesting that it may be possible to violate the WFL at
a QCP.\cite{Senthil0305193,Podolsky2005}

Experimentally, a verification of the WFL was observed in
CeNi$_2$Ge$_2$.\cite{KambeJLT99} While this study was a singular
measurement at zero magnetic field and ambient pressure, the
belief is that this point in phase space lies in close proximity
to an antiferromagnetic QCP. Recently, the WFL was also confirmed
in the field tuned quantum critical system
CeCoIn$_5$.\cite{PaglioneWFL2006} Another possibly relevant system
in which the WFL has been measured are the high T$_c$
cuprates.\cite{HillNature, Proust2005} In this case a violation
has been observed in the field-induced normal state, which may be
related to an underlying QCP in the phase diagram.

In this letter, we have chosen to study the WFL in \Sr327, a
bilayer perovskite material.\cite{PerryPRL} The well-known single
layer compound Sr$_2$RuO$_4$ is believed to be a spin triplet
superconductor\cite{MackenzieRMP}, while the infinite layer
compound SrRuO$_3$ is an itinerant electron
ferromagnet\cite{Callaghan1966}. When the magnetic field is
applied in the RuO$_2$ planes of \Sr327,~a first order
metamagnetic transition is observed. The critical endpoint of this
line of first order metamagnetic transitions is systematically
driven to zero temperature at $H$~=~7.85~T by rotating the
magnetic field out of the plane\cite{GrigeraPRB03}. Thus, the term
quantum critical endpoint(QCEP) is used when discussing \Sr327.
The use of magnetic field as a tuning parameter in this
stoichiometric system allows for a very sensitive test of the
Wiedemann-Franz law to be made on a clean system. We present
thermal and charge conductivity data on \Sr327, which demonstrate
that the integrity of the electron does survive in the vicinity of
a QCEP as the WFL is satisfied at all fields.

A second aspect of this study is that finite temperature thermal
and charge conductivity data allow us to comment on the nature of
the fluctuations present in this system. While the above
circumstantial evidence would suggest that ferromagnetic
fluctuations are the most relevant magnetic excitations here,
neutron scattering finds both antiferromagnetic and ferromagnetic
fluctuations of similar strength in zero
field,\cite{CapognaPRB2003} thus leaving an open question as to
what fluctuations may be relevant for this metamagnetic QCEP. We
find that the temperature dependence of the Lorenz ratio is in
good agreement with the standard picture of itinerant electron
metamagnetism, which thus implies that the relevant fluctuations
are ferromagnetic in nature.

Thermal conductivity was measured using a two thermometer one
heater setup described elsewhere\cite{Kappasetup}. The absolute
accuracy is limited by the uncertainty in the geometric factor of
the sample ($\sim$ 10 \%), but the relative changes between
different fields is limited by the accuracy of the thermometer
calibration which is $\sim$ 1\% for the temperature sweeps, and
due to the magnetoresistance of the thermometers can drift to as
much as 5\% for the field sweeps. The \Sr327~single crystals
studied were grown in a floating zone furnace\cite{Perrygrowth}.
With the exception of the data in figure 2(b)(which has a residual
resistivity of 0.5 $\mu\Omega$~cm and was aligned
$\sim$20$^{\circ}$ off the c-axis) data is presented for a sample
with a residual resistivity of 2.1 $\mu\Omega$~cm, and was aligned
with the field parallel to within 5$^{\circ}$ of the c-axis. The
transport was in the {\it ab}-plane.

\begin{figure}[t]
\centering \leavevmode \epsfxsize=8.5cm
\epsfbox{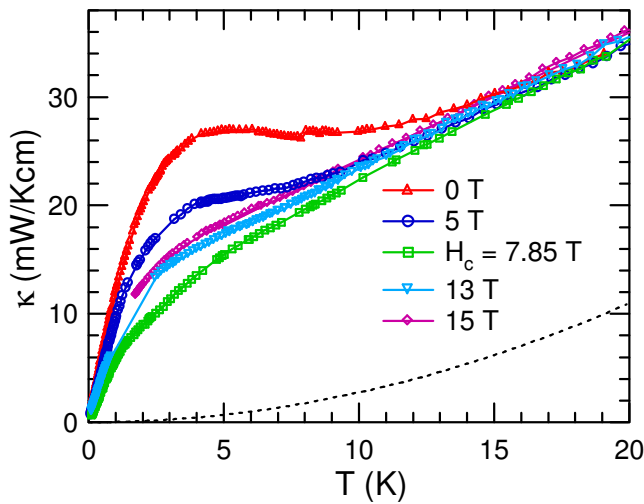} \vspace{.2cm} \caption{(color
online) High temperature thermal conductivity at several fields
along the {\it c}-axis including the critical field. The heat
current was applied in the {\it ab}-plane. The dashed line is an
estimate of the phonon conductivity obtained by considering the
WFL as discussed later in the text.}\label{fig1}
\end{figure}

Figure 1 presents the raw thermal conductivity data at, above, and
below the critical field of $H_c$ = 7.85 T. In zero field, the
thermal conductivity has a peak at approximately 5~K which
vanishes as one reaches the critical field. At still higher
fields, the peak begins to return, albeit much more slowly.
Comparison with the limited specific heat data on single crystals
shows a qualitatively similar behavior
\cite{PerryPRL,ZhouPRB2004}. $C/T$ has a peak at zero field, which
is heavily suppressed close to the critical field. At this point
the specific heat begins to diverge, indicative of non-Fermi
liquid behavior. Theoretical work is needed to evaluate how the
changing electronic structure affects both the number of carriers
and the scattering rate, both of which enter into the thermal
conductivity.

\begin{figure}[t]
\centering \leavevmode \epsfxsize=8.5cm
\epsfbox{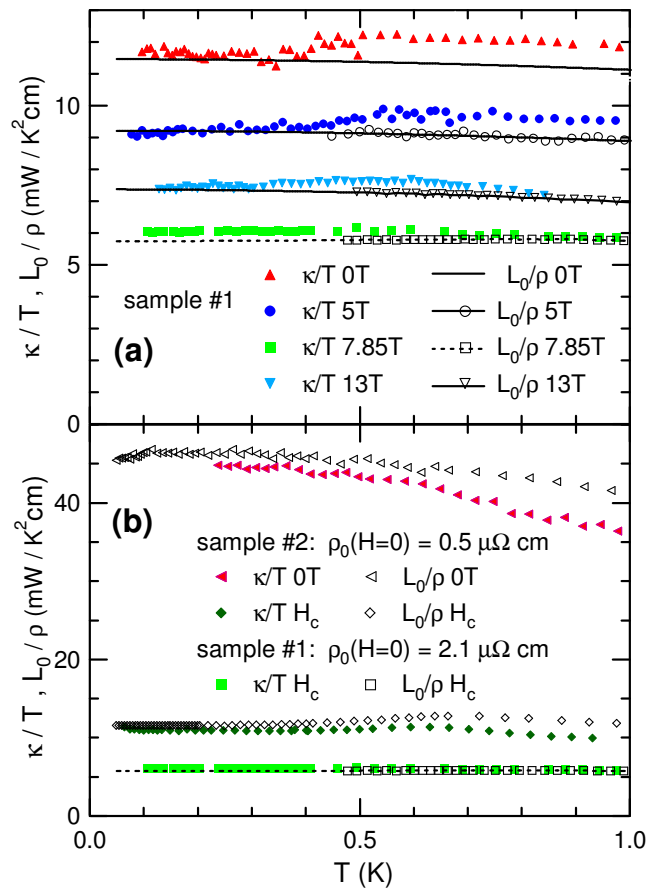} \vspace{.2cm} \caption{(color
online) A comparison of heat ($\kappa/T$) and charge
($L_0$/$\rho$) conductivities at several fields versus $T$ at low
temperature. These quantities extrapolate to the same value at $T$
= 0 if the Wiedemann-Franz law is to be satisfied. Solid lines are
fits to $\rho = \rho_0 + AT^2$. The dashed line for $H_c$ is a
linear fit below 0.8~K.(b) Comparing two samples with different
amounts of disorder. Sample \#2 has been rotated by 20$^{\circ}$
with respect to the c-axis which mainly changes the value of the
critical field.}\label{fig2}
\end{figure}

We now turn our attention to the low temperature end of the data.
By plotting $\kappa/T$ vs $T$ the intercept represents the
fermionic contribution to the thermal conductivity resulting from
a constant density of states in the limit that scattering is
dominated by elastic scattering. From the resistivity data
(plotted as $L_0$/$\rho$ for comparison with the WFL) we can see
in figure 2(a) that we have clearly reached that limit below 1~K.
As there is almost no temperature dependence we can reliably
extrapolate our results to $T$ = 0, where we find that the WFL
(which states that $\kappa$/$\sigma~T~=~L_0$) is satisfied to
within 5\% at all fields.The phonon conductivity does not have a
zero temperature intercept in $\kappa/T$ and hence does not enter
into the discussion at this time.

\begin{figure}[t]
\centering \leavevmode \epsfxsize=8.5cm
\epsfbox{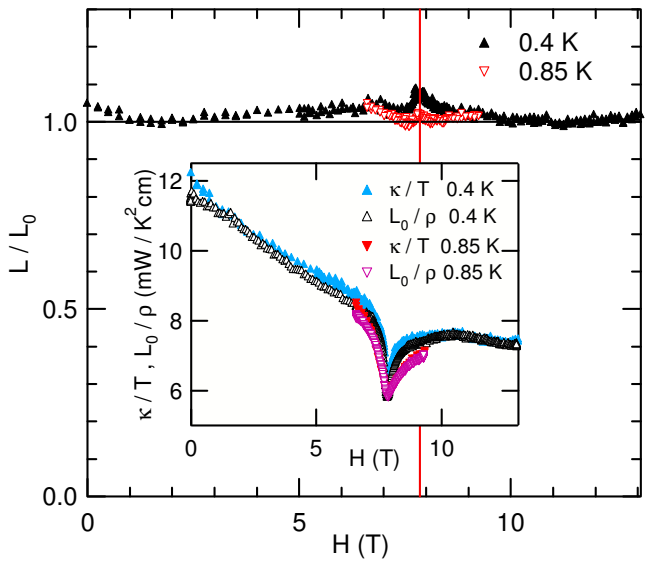} \vspace{.2cm} \caption{(color
online) Verification of the Wiedemann-Franz law from field sweeps
of the Lorenz number. Inset: Field sweeps of heat ($\kappa/T$) and
charge ($L_0/\rho$) conductivity at low temperature.}\label{fig3}
\end{figure}

The lack of temperature dependence at all fields also allows us to
comment on the elastic scattering from finite temperature field
sweeps. As a result, in the inset to figure 3 we plot $\kappa/T$
and $L_0$/$\rho$ as a function of field at low temperatures. The
lowest temperature curves are clearly a good approximation to the
$T$=0 value for the residual elastic scattering. The resistivity
shows a pronounced peak (dip in $L_0$/$\rho$) at the critical
field, the origin of which is unknown.\cite{GrigeraScience}
Clearly, however, the thermal conductivity data tracks this dip.
To see how precisely the thermal conductivity tracks the charge
conductivity we plot the Lorenz ratio
($L/L_0$=$\kappa$/$\sigma$$TL_0$) in the main panel of figure 3.
The fact that the WFL is satisfied is again demonstrated by the
fact that this quantity equals 1 at all fields. The slow drift in
the value of the Lorenz number is due to calibrating the
magnetoresistance of the thermometers used to measure the
temperature gradient in the sample. The sharper bump at the
critical field likely results from the new phase which emerges to
"protect" the QCEP \cite{PerryPRL2004}. The slight enhancement
observed in $L$ at finite temperature in this protected phase may
result from either a reduction in inelastic scattering or a shift
to large angle scattering upon entering this new phase.

As disorder can dramatically modify the behavior at a QCP, we
investigate its effects by also measuring a higher purity
crystal\cite{PerryPRL2004} shown in figure 2(b). This is
particularly true of the phase which protects the QCEP. In zero
field one can observe the factor of 4 increase in purity relative
to the sample presented in figure 2(a), although at the critical
field the change in elastic scattering is not nearly as great. The
field is also applied slightly off the c-axis which has the added
advantage that the resistive anomaly associated with the phase
which protects the QCEP is enhanced. This anomaly is mimicked in
the thermal data and at $T$=0 we find that this phase also obeys
the WFL to within 5\%.

The presence of exotic fermionic excitations which do not carry
charge $\pm$$e$ (such as charge 0 spinons) or charged bosonic
excitations would result in a violation of the WFL. The
verification of the WFL in the $T\rightarrow$~0~K limit proves
that such excitations do not exist here.

For charge $\pm$$e$ quasiparticles, the WFL is valid only when
heat and charge transport are affected equally, as is the case for
elastic processes dominant in the $T\rightarrow$~0~K limit. As
temperature is increased, inelastic scattering dominates the
elastic scattering of quasiparticles. Thus, a finite temperature
study of the Lorenz number may also shed light on to the nature of
the excitations relevant to the non-Fermi liquid behavior observed
at the critical field. The strength of such a study was
exemplified in CeRhIn$_5$ where thermal measurements found
quantitative agreement with the antiferromagnetic spin fluctuation
spectrum measured by inelastic neutron
scattering.\cite{PaglioneRh115PRL2005}

\begin{figure}[t]
\centering \leavevmode \epsfxsize=8.5cm
\epsfbox{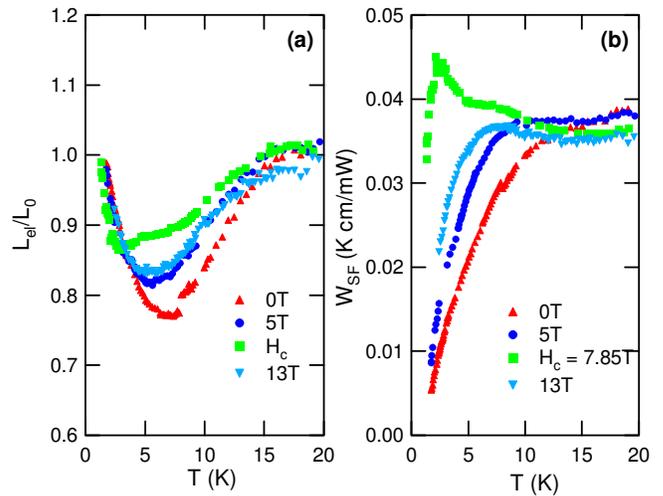} \vspace{.2cm} \caption{(color
online) (a) Temperature dependence of the electronic contribution
to the Lorenz ratio $L(T)$, which is consistent with the standard
picture for itinerant electron metamagnetism. (b) The thermal
resistance $W_{SF} \equiv 1/\kappa_{el} - \rho_0/L_0T$ attributed
to scattering off of spin fluctuations.}\label{fig4}
\end{figure}

Figure 4(a) shows the temperature dependence at a few fields
including the critical field. A phonon conductivity term of
$\kappa_{ph} = 0.03 (mW/K^3cm) T^2$ independent of field for
phonons being scattered by electrons was subtracted from the
thermal conductivity data, to give purely the electronic
contribution to $L$. This choice of phonon subtraction is the
minimum necessary to place $L_{el}$ in the physically allowable
parameter range $0\leq L_{el}\leq L_{0}$, but the conclusions
drawn from these data are insensitive to the amount of phonon
conductivity assumed. Interestingly, the effect of small angle
inelastic scattering becomes stronger as one approaches the
critical field as signified by the increasingly rapid suppression
of $L$ at low temperatures. This is consistent with the divergence
of the quasiparticle-quasiparticle scattering at $H_c$, as deduced
from the $T^2$ coefficient of
resistivity.\cite{GrigeraScience,PaglioneWFL2006} As a result, the
minimum in $L(T)$ naturally also moves to lower temperature. The
puzzling observation is that the magnitude of the dip in $L(T)$ is
gradually suppressed as the critical field is approached. This is
partially a result of the increase in elastic scattering about the
critical field. However, the effect is still present if the
elastic scattering is removed $L_{inel} = (\kappa/T -
\kappa/T(T=0))(\rho - \rho(T=0)$ (not shown). In CeRu$_2$Si$_2$, a
heavy fermion metamagnetic system, the $T\rightarrow 0$ limit was
not explored, but at finite temperature the behavior in $L(T)$ was
similar to that observed here.\cite{SeraPRB1997}

Theoretically, heat and charge transport properties have been
calculated for a nearly ferromagnetic metal.\cite{UedaandMoriya}
 It was found that as the Stoner parameter
$\alpha$ is enhanced the deviation of the Lorenz number from the
Sommerfeld value $L_0$ is reduced. Thus, in this picture our data
would indicate that the Stoner parameter increases as the critical
field is approached. This is precisely what happens in the
standard picture for itinerant electron
metamagnetism.\cite{Wohlfartand Rhodes} By applying a magnetic
field, the spin-up and spin-down bands are pushed in opposite
directions. When a peak in the density of states exists near the
Fermi level, then a metamagnetic transition will occur when the
peak in one of the spin-split bands is pushed through the chemical
potential. The Stoner parameter $\alpha$, which is equal to the
density of states at the chemical potential times the interaction
term $U$, is thus maximal at the metamagnetic transition, which is
in agreement with our field dependence of $L(T)$. Furthermore, the
thermal resistance due to spin fluctuations $W_{SF} \equiv
1/\kappa_{el} - \rho_0/L_0T$ shown in figure 4(b) strongly
resembles that of the calculations done by Ueda and
Moriya,\cite{UedaandMoriya} further emphasizing the point that the
behavior in $L(T)$ is not solely due to a change in the elastic
scattering.

This simple model of metamagnetism remarkably reproduces many
aspects of the observations seen in \Sr327 in addition to the
behavior in the Lorenz number. It is natural to anticipate a peak
in the density of states from a Van-Hove singularity in this
two-dimensional system. Fine tuning of U and the initial form of
the density of states in a Stoner model as described can provide a
fair explanation of the magnetization\cite{BinzandSigrist} and
specific heat data\cite{RonningandGrafunpublished}. In addition,
it has been conjectured that the unidentified phase protecting the
QCEP may be a result of a type of Pomeranchuk instability of the
Fermi surface,\cite{KeeandKim,GrigeraScience2004} or alternatively
from nanoscale charge inhomogeneity,\cite{Honerkamp2005} as the
van-Hove singularity passes through the chemical potential. In
fact, the van-Hove singularity is one possible microscopic origin
for the effective action used by Millis \textit{et
al.}\cite{MillisPRL2002} It should be pointed out that the
renormalization group treatment of this action has proven quite
successful in explaining the thermal expansion
data\cite{gegenwartSr327} among other experimental observations.

In conclusion we have observed the verification of the
Wiedemann-Franz law in \Sr327~in the limit as $T\rightarrow0$ at
all fields including the field at which a quantum critical
end-point occurs, which implies that there is no breakdown of the
electron at the QCEP. More precisely, there are no additional
fermionic carriers of heat (such as spinons) other than those
which carry charge $e$. This further supports the notion that the
QCEP in \Sr327 can be described in the Hertz-Millis formalism for
quantum criticality. It will be interesting to see if the WFL is
still satisfied in other quantum critical systems in which the
Hertz-Millis theory fails. Finally, the finite temperature data
are consistent with the standard picture for itinerant electron
metamagnetism, and, as a result, one should expect that the
ferromagnetic fluctuations are responsible for the observed
non-Fermi liquid behavior at the quantum critical endpoint.

We thank Y.B. Kim, A. Schofield, and I. Vekhter for useful
discussions. Work at Los Alamos was performed under the auspices
of the US DOE.


\end{document}